\documentclass[secnumarabic, pra, showkeys, showpacs]{revtex4}
\usepackage{bm}
\usepackage{hyperref}
\usepackage{graphicx}
\usepackage{amsmath}

\begin{document}

\title{Absolute motion determined from Michelson-type experiments in optical media}
\author{Valery P. Dmitriyev}
\affiliation{Lomonosov University, Moscow, Russia}
\email{dmitriyev@cc.nifhi.ac.ru}
\date{August 25, 2010}

\begin{abstract}
The symmetry of vacuum is characterized by the Lorentz group with the parameter $c$. Physical space inside the homogeneous optical medium should be described by the Lorentz group with the parameter $c/n$, where $n$ is the refractive index of the medium. Violation of a one-parameter phenomenological symmetry in the discrete medium, such as gas, creates the opportunity for the experimental detecting the motion of the optical medium relative to luminiferous aether.
\end{abstract}
\maketitle

The anisotropy of the speed of wave propagating in the otherwise isotropic medium arises when the medium moves in the laboratory reference frame.  We are interested in a case when at least the part of the anisotropy is retained in the reference frame moving together with the medium.

\section{Conditions for anisotropy of wave speed in a moving medium}

 The linear-elastic incompressible medium is governed by d'Alembert equation
 \begin{equation}
\partial^2_t\mathbf{s}=w^2\nabla^2\mathbf{s}\label{dalambert equation}
\end{equation}
where $\mathbf{s}$ is the variable (displacement) and $w$ the elastic constant of the medium.
 Equation ({\ref{dalambert equation}) is Lorentz-invariant. Let a \textit{signal} move through the medium with the velocity $u<w$. Assuming that inertial reference frames are related with each other by the Lorentz transformation we have the following group transformation of $u$ to the moving with the velocity $v$ reference frame
\begin{equation}
u'=\frac{u-v}{1-uv/w^2}.\label{velocity transformation}
\end{equation}

Taking in (\ref{velocity transformation}) $v\rightarrow-v$ and then $u'\rightarrow\tilde{u}$, we may interpret $\tilde{u}$ as the result of the drag of the signal by the moving with the velocity $v$ medium:
\begin{equation}
\tilde{u}=\frac{u+v}{1+uv/w^2}.\label{drag}
\end{equation}
Passing with (\ref{velocity transformation}) into the reference frame of the medium, we will obviously find $\tilde{u'}=u$, i.e. the speed of the signal becomes isotropic again. Expanding (\ref{drag}}) over $v\ll w$ gives
\begin{equation}
\tilde{u}=u+v(1-u^2/w^2)-uv^2/w^2(1-u^2/w^2)+...\,\,.\label{expansion of drag}
\end{equation}
Dropping in (\ref{expansion of drag}) higher orders, we will obtain the model
\begin{equation}
\tilde{u}_\textrm{F}=u+v(1-u^2/w^2)\label{Fresnel drag}
\end{equation}
referred to as the Fresnel drag.
Substituting $\tilde{u}_\textrm{F}$ for $u$ in the right-hand part of (\ref{velocity transformation}) gives the anisotropy of the signal's speed as observed in the moving together with the medium reference frame:
\begin{equation}
\tilde{u}'_\textrm{F}\approx u\left[1+\frac{v^2}{w^2}\left(1-\frac{u^2}{w^2}\right)\right],\label{anisotropy}
\end{equation}
i.e. the signal's speed in the direction collinear to $\mathbf{v}$ will be (\ref{anisotropy}) while in the perpendicular direction $u$ as before. Still, this anisotropy is of the same order as the terms dropped out in (\ref{expansion of drag}).

We may specify (\ref{dalambert equation}) as $w=c/n$  with $c$ the speed of light in vacuum and $n$ refractive index of a dielectric medium. Take notice in (\ref{anisotropy}) that $\tilde{u}'_\textrm{F}=u$ when $u=w$. This is realized in two cases:

1) $w=c$ and $u=c$, i.e. in vacuum;

2) $n>1$, $w=c/n$ and $u=c/n$, that corresponds to a dense homogeneous optical medium (cf. \cite{Wagner}).\\
Supposedly, in a discrete medium, such as rarified gas ($n\sim1$), we will have $w=c$ whereas $u=c/n$, and then $\tilde{u}'_\textrm{F}\ne u$. Therefore, the anisotropy of wave propagation in the moving together with the medium reference frame may arise provided that this "medium" represents a small or point-like perturbation of vacuum.

\section{Michelson interferometer in dielectric media}

 Now we may calculate for $w=c$, $u=c/n$ the difference of wave's round-trip times in the collinear and transverse to $\mathbf{v}$ orientations of the Michelson interferometer. The collinear path time can be immediately found from (\ref{anisotropy}) as $t'_\parallel=2l/\tilde{u}'_\textrm{F}$, where $l$ is the arm's length of the interferometer. However, $t'_\perp\ne2l/u$ since in the transverse direction, because of the inclination of the wave's trajectory, there should also be taken into account the drag of the wave by the medium. So, it is convenient to perform calculations in the absolute reference frame (see Appendix). Thus we obtain for $v\ll c$
 \begin{equation}
\Delta t'\approx\frac{v^2}{c^2}\frac{l}{cn}(n^2-1)(2-n^2)\label{time difference}
\end{equation}
(here we have taken $\Delta t'\approx\Delta t$ since by the dilation of time $\Delta t'=\Delta t\sqrt{1-v^2/c^2}$).
By the initial supposition, (\ref{time difference}) is valid only for $n\gtrsim1$. Nevertheless, formula (\ref{time difference}) is capable to describe, at least in general features, the dependence of $\Delta t'$ on the refractive index of optically denser media.

In agreement with the conditions above stated, experiments on gas-mode Michelson interferometer ($n_\textrm{air}\approx1.0003$) indicate the anisotropy of light speed, estimating by means of (\ref{time difference}) the velocity $v$ of the Earth in aether as several hundreds kilometers in a second \cite{Cahill, Demjanov second order}. While experiments in the solid optical monolith ($n=1.5\div1.75$) show no interference fringe shift \cite{ShamirFox, Cahill book} or an equivalent \cite{Nagel}, i.e. $\Delta t'_{\textrm{solid}}=0$. This agrees with $w=c/n,\,u=c/n$ case. We may expect that for moderate optical densities an intermediate between $w=c,\,u=c/n$ and $w=c/n,\,u=c/n$  phenomenology takes place. So that $\Delta t'$ first grows with the increase of $n$, and further declines to null. This corresponds to behavior of the function (\ref{time difference}) in the range $1<n^2\leq2$. Still, Demjanov \cite{Demjanov second order} claims that the experimental curve measured by him changes the sign and goes to negative in accord with  (\ref{time difference}) at $n^2\geq2$.

\section {Conclusion}

The crucial point in detecting the anisotropy of light propagation in the moving together with the medium laboratory is the validity of the Fresnel drag (\ref{Fresnel drag}) in the absolute reference frame, i.e. in aether. In this event, it is important that the Fresnel drag of light, obtained above from kinematical relation (\ref{drag}), can be derived independently in optics of moving media \cite{Drezet}.  One way or another, the accuracy of the derivation must be higher than $v^2/w^2$.

\appendix

\section{}

We consider Michelson interferometer whose working chamber is filled with the dielectric medium that is at rest in the device. Supposedly the interferometer moves (along with the Earth) uniformly with a velocity $v>0$ in the absolute reference frame, where $v\ll c$.

To account for the drag of light by the moving medium, Fresnel formula with the appropriate sign before $v$ can be used
\begin{equation}
\tilde{c}_\pm\approx\frac{c}{n}\pm v\left(1-\frac{1}{n^2}\right).\label{Fresnel}
\end{equation}
 In order to compute the light round-trip time  $t_\parallel$ in collinear to $\mathbf{v}$ orientation, we will use classical addition $\tilde{c}_\pm\mp v$ with subsequent account of the Lorentz contraction \cite{ShamirFox}.  This gives in the absolute reference frame
\begin{equation}
t_\parallel=\frac{l_\parallel}{\tilde{c}_+-v}+\frac{l_\parallel}{\tilde{c}_-+v}=l\sqrt{1-\frac{v^2}{c^2}}\left(
\frac{1}{\cfrac{c}{n}-\cfrac{v}{n^2}}+\frac{1}{\cfrac{c}{n}+\cfrac{v}{n^2}}\right)\approx
\frac{2l}{c}n\left[1+\frac{v^2}{c^2}\left(\frac{1}{n^2}-\frac{1}{2}\right)\right]\label{time parallel}
\end{equation}
where the longitudinal arm of the interferometer $l_\parallel=l\sqrt{1-v^2/c^2}$, and terms of the order higher than $v^2/c^2$ were neglected.
\begin{figure}[h]
\begin{picture}(200,65)(0,0)
\put(80,0){\vector(0,1){60}} \put(80,60){\vector(1,0){30}} \put(80,0){\vector(1,2){30}}
\put(81.7,18){$\alpha$} \put(93,63){$v$} \put(102,30){$\dfrac{c}{n}+v\sin\alpha\left(1-\dfrac{1}{n^2}\right)$}
\put(57,30){$2l/t_\perp$}
\end{picture}
\caption{Velocity triangle in the transverse arm of the medium-filled Michelson interferometer.}\label{fig2}
\end{figure}
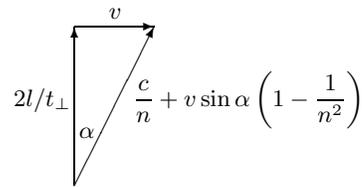

Accounting for tangential incline of light in the transverse arm, we will consider the velocity triangle (Fig.\ref{fig2}). Here the light is dragged due to projection of $\mathbf{v}$ on $\mathbf{c}/n$. Insofar as the inclination angle $\alpha$ is small, it can be approximated by
\begin{equation}
\sin\alpha=\frac{v}{c/n+v\sin\alpha(1-1/n^2)}\approx\frac{vn}{c}.\label{inclination angle}
\end{equation}
Then we have from Fig.\ref{fig2}
\begin{equation}
l^2+\left(\frac{vt_\perp}{2}\right)^2\approx\left[\frac{c}{n}+\frac{v^2n}{c}\left(1-\frac{1}{n^2}\right)\right]^2\left(\frac{t_\perp}{2}\right)^2.\label{velocity triangle}
\end{equation}
Relation (\ref{velocity triangle}) gives for the transverse round-trip time $t_\perp$ in the reference frame of aether: \begin{equation}
t_\perp\approx\frac{2l}{\sqrt{\left[\cfrac{c}{n}+\cfrac{v^2n}{c}\left(1-\cfrac{1}{n^2}\right)\right]^2-v^2}}\approx \frac{2ln}{c}\left[1-\frac{v^2}{c^2}\left(\frac{n^2}{2}-1\right)\right]\label{time transverse}
\end{equation}
where terms of the order higher than $v^2/c^2$ were neglected. Formula (\ref{time transverse}) differs drastically from that obtained for $t_\perp$ in \cite{ShamirFox}.

Subtracting (\ref{time parallel}) from (\ref{time transverse}) we obtain
\begin{equation}
\Delta t=t_\perp-t_\parallel\approx \frac{v^2}{c^2}\frac{l}{cn}(n^2-1)(2-n^2)=\frac{v^2}{c^2}\frac{l}{c\sqrt{\varepsilon}}\Delta \varepsilon(1-\Delta \varepsilon)\label{time difference Demjanov}
\end{equation}
where $\Delta\varepsilon=n^2-1$ accounts for the contribution of the particles of matter into the dielectric permittivity $\varepsilon=n^2$ of the luminiferous medium, aether plus the dielectric substance. Formula (\ref{time difference Demjanov}) has been first proposed by Demjanov in order to describe the run of the experimental curve obtained by him from measurements on Michelson interferometer in various optical media and at different wavelengths \cite{Demjanov second order}.

\end{document}